\documentstyle[aps,preprint,psfig,eqsecnum]{revtex}
\tightenlines 
\begin{document}
\draft 
\title{Testing Scalar-Tensor Gravity Using Space Gravitational-Wave
Interferometers}
 
\author{Paul D. Scharre and Clifford M. Will \cite{cmw}}       

\address{McDonnell Center for the Space Sciences, Department of Physics, \\Washington University, 
St. Louis, Missouri
63130}

\maketitle   

\begin{abstract}
We calculate the bounds which could be placed on scalar-tensor theories of
gravity of the Jordan, Fierz, Brans and Dicke type 
by measurements of gravitational waveforms from neutron stars
(NS) spiralling into massive black holes (MBH) using LISA, the proposed
space laser interferometric observatory.  Such observations 
may yield significantly more stringent bounds on
the Brans-Dicke coupling parameter $\omega$ than are achievable from
solar system or binary pulsar measurements.  For NS-MBH inspirals,
dipole gravitational radiation modifies the inspiral and generates an
additional contribution to the phase evolution of the emitted gravitational
waveform.  Bounds on $\omega$ can therefore be found by using the
technique of matched filtering.  We compute the Fisher information matrix
for a waveform accurate to second post-Newtonian order, including the
effect of dipole radiation, filtered using a currently modeled noise curve
for LISA, 
and determine the bounds on $\omega$ for
several different NS-MBH canonical systems.  For example, observations of
a $1.4 M_{\odot}$ NS inspiralling to a $10^3 M_{\odot}$ MBH with a
signal-to-noise ratio of 10 could yield a
bound of
$\omega
> 240,000$, substantially greater than the current experimental bound of
$\omega > 3000$.
\end{abstract}

\section{Introduction and Summary}     

The observation of gravitational waves from inspiralling compact binaries is expected to provide
an excellent testbed for theories of gravity alternative to
general relativity (GR) (see \cite{phystoday} for a review).  
In earlier work  \cite{CW1}, we have shown that
the measurement of gravitational waves from neutron star 
(NS) - black hole (BH) mergers by ground-based detectors such as the Laser
Interferometric Gravitational-Wave Observatory (LIGO) could bound 
the strength of a scalar field in scalar-tensor theories of
gravity.  The bounds estimated in \cite{CW1} 
were potentially better than the solar system
bounds that were in force at the time (1994), 
but are no longer competitive with bounds
that have improved through the use of Very Long Baseline Interferometry
(VLBI) to measure the deflection of light by the sun.

In this paper, we analyze potential bounds on scalar-tensor theory
made possible by plans for space-based gravitational-wave 
observatories such as the 
Laser Interferometer Space Antenna (LISA), a proposed project 
of the National Aeronautics and 
Space Administration (NASA) and the European Space Agency (ESA)
\cite{danzmann}.  
We show that the detection 
of NS inspirals into massive black
holes (MBH) by LISA
may provide an opportunity to place 
significantly more stringent bounds on scalar-tensor theory than 
current solar-system or binary-pulsar bounds or those achievable
by an
Earth-based detector.

Space-based interferometers are being designed to detect gravitational-wave 
signals
in a lower frequency band than ground-based detectors such as
LIGO, GEO, VIRGO, and
TAMA, which measure waves from approximately 10 Hz to $10^4$ Hz.  The latter
interferometers are expected to detect waves from low-mass inspiralling
compact binaries (masses $1-10 M_{\odot}$).  LISA is being designed to detect
inspirals and mergers
of massive black holes (MBH) with masses of $10^3 - 10^6
M_{\odot}$ , among other potential sources \cite{danzmann}.  
Mergers of such high-mass objects will produce
waves in LISA's proposed sensitive band 
of frequencies between $10^{-4}$ Hz and  $10^{-1}$ Hz.  

The inspiral of such binary systems is driven by gravitational
radiation reaction, which imprints onto the phase and amplitude of the
emitted gravitational waves information about the masses and spins of
the orbiting bodies, about the orbital elements and their evolution, and about
the
theory of gravity that is in force.  Because broad-band
interferometric detectors are particularly sensitive to the phasing of
the waves, accurate determinations of system and theoretical
parameters are possible \cite{3min}.  

The simplest version of scalar-tensor gravity is that studied by 
Jordan, Fierz, Brans and 
Dicke, commonly known as Brans-Dicke (BD) theory \cite{BD1}.  In
BD, a scalar gravitational field $\phi$ is postulated in addition to 
the spacetime metric $g_{\mu\nu}$,  
with an effective strength
inversely proportional to a coupling parameter $\omega$.  In BD, $\omega$ 
is 
assumed to be a fixed constant.  (Generalized scalar-tensor theories that are 
currently popular 
byproducts of string theory and other unification schemes treat $\omega$ 
as a function of $\phi$.)  
As $\omega \to \infty$, the effect of the
scalar field tends to zero and BD $\to$ GR.  Current
experimental bounds on $\omega$ from measurements of light bending place
$\omega > 3000$ (for a review see \cite{Willlivrev}; for recent light
bending measurements, see \cite{eubanks}).

The details of the behavior of compact binaries in scalar-tensor
gravity have been
worked out by Eardley \cite{Eardley75}, Will \cite{Will77}, Zaglauer
\cite{WillZaglauer}
and Damour and Esposito-Far\`ese 
\cite{damouresposito}.  
The chief difference between general
relativity and scalar-tensor gravity is the existence, 
in the latter, of dipole gravitational radiation.  
In general relativity, for slowly moving systems, the leading
multipolar contribution to gravitational radiation is quadrupolar, with
the result that the dominant radiation-reaction effects are at order
$(v/c)^5$ relative to Newtonian gravity, where $v$ is the orbital
velocity.  
The rate, due to quadrupole radiation in GR, at which a binary system loses
energy is given 
by
\begin{equation}
\left(\frac{dE}{dt}\right)_{\rm quadrupole} = -\frac{8}{15} 
\frac{\eta^2 m^4}{r^4}
(12 v^2-11 \dot{r}^2) \,,
\label{Edotquad}
\end{equation}
where $r$, $v$, and $\dot{r}$ represent the
orbital
separation, relative orbital velocity, and radial velocity, respectively.  
We use units in which $G=c=1$.  
The 
quantities $\eta$ and $m$ are the reduced mass parameter and 
total mass, respectively, given by 
$\eta = m_1m_2/(m_1+m_2)^2$, 
and $m = m_1 + m_2$.  In BD, this formula is modified by corrections
to the coefficients of $O(1/\omega)$
(BD also predicts monopole radiation, but in binary systems it 
contributes only to these $O(1/\omega)$ corrections). 

The 
important modification in BD is the additional energy loss
caused by dipole radiation.
By analogy with electrodynamics, dipole radiation is a
$(v/c)^3$ effect, potentially much stronger than quadrupole radiation.
However, in scalar-tensor theories, the gravitational ``dipole
moment'' is governed by the difference $s_1 - s_2$ between the bodies,
where $s_i$ is a measure of the self-gravitational binding energy per
unit rest mass of
each body.  Technically, $s_i$ is the ``sensitivity'' of the total mass
of the body to variations in the background value of Newton's
constant, which, in these theories, is a function of the scalar
field:  
\begin{equation}
s_i = \left( \frac{\partial (\mbox{ln }m_i)}{ \partial (\mbox{ln }G) }
\right)_N \,,
\label{sensitivity}
\end{equation}
where $G$ is the effective Newtonian constant at the star and the
subscript
$N$ denotes holding baryon number fixed.  
The energy loss caused  by dipole radiation is given by
\begin{equation}
\left(\frac{dE}{dt}\right)_{\rm dipole} = -\frac{2}{3} \frac{\eta^2
m^4}{r^4}\left(\frac{{\cal S}^2}{\omega}\right) \,,
\label{Edotdipole}
\end{equation}
valid to first order in $1/\omega$, where 
${\cal S} \equiv s_1 - s_2$.

In BD, the
sensitivity of a black hole is always $s_{BH} = 0.5$, while the sensitivity
of a neutron star varies with the equation of state and mass.  For
example,  $s_{NS} \approx 0.12$ for a $1.4 M_{\odot}$ NS with a stiff equation
of state, and $s_{NS} \approx 0.19$ for a $1.1 M_{\odot}$ NS with a soft
equation of state.  A more detailed discussion of neutron 
star sensitivities can be
found in \cite{WillZaglauer,Zaglauer} in the context of 
Brans-Dicke theory, and in \cite{damouresposito} in the context of 
general scalar-tensor gravity.  For white dwarfs, $s_{WD} < 10^{-3}$.

Ironically, binary black-hole systems are not at all promising
for studying dipole gravitational radiation because $s_{BH1} - s_{BH2} \equiv
0$, a consequence of the no-hair theorems for black holes.
Essentially they radiate away any scalar field, so that a binary black
hole system in scalar-tensor theory behaves as if GR were valid (see
\cite{CW1} for further discussion).
Similarly, binary neutron star systems, such as the Hulse-Taylor
binary pulsar PSR 1913+16 and similar systems, are also not effective
testing grounds for dipole radiation \cite{WillZaglauer}.  
This is because neutron star
masses tend to cluster around the Chandrasekhar mass of $1.4 M_\odot$,
and the sensitivity of neutron stars is not a strong function of mass
for a given equation of state.
Hence in systems like the binary pulsar, dipole radiation is naturally
suppressed by symmetry, and the bound achievable cannot compete with those from
the solar system \cite{caveat}.  
Thus the most promising systems are mixed: BH-NS,
BH-WD, or NS-WD.

While a leading scientific motivation for LISA is detection of waves
from the merger of binary massive black holes in the centers of
galaxies at cosmological distances, LISA will also be able to study
waves from the inspiral of low-mass compact objects (black holes,
neutron stars, white dwarfs) into massive black holes in galaxies and
globular clusters.  The orbits of these inspiralling bodies are
expected to be complex: possibly highly eccentric because the bodies
will have been sent toward the hole by gravitational scattering from
stars in the surrounding cluster; possibly strongly perturbed
by the surrounding stars each time the body passes through its
apholion; possibly dragged in a precessing orbit by the Lens-Thirring
effect of a rapidly rotating hole.  Analysing the gravitational waves
from such orbits will be a challenging task.

However, to study the potential for testing scalar-tensor gravity, we
will make a drastic simplification of the problem: we will assume that
the compact body spirals toward a non-rotating hole on an
adiabatically shrinking circular orbit, up to the innermost stable
circular orbit (ISCO) of the black hole, whereupon the body plunges
across the horizon on a short timescale.  We will assume that the
waves from such a
quasi-circular orbit can be studied for up to a year using LISA.

With these assumptions, one can make a crude estimate of the bounds on
scalar-tensor theory that could be possible.  
The idea is to use matched filtering of a theoretical template waveform
against
the output of the detector.  A ``match'' is a signal whose total
accumulated phase over the integration time of the experiment
matches that of the template 
within a number of radians set by the time resolution of the
instrument and the relevant frequency.
The accumulated phase is
given by
\begin{equation}
\Phi_{\rm GW} = \int_{t_{\rm i}}^{t_{\rm f}} 2\pi f dt
	= \int_{f_{\rm i}}^{f_{\rm f}} 2\pi (f/ \dot f )  df \,,
\label{phase}
\end{equation}
where $f$ is the gravitational-wave frequency (twice the orbital
frequency),
and the subscripts ``i'' and ``f'' denote the values at the beginning
and end of the integration.  The evolution of the frequency is given
by
\begin{equation}
\dot u = {\cal M}^{-1} (96/5) u^{11/3} (1+ b \eta^{2/5} u^{-2/3} )
\,,
\label{freqevol}
\end{equation}
where ${\cal M} \equiv \eta^{3/5} m$ is the ``chirp'' mass, 
and $u=\pi {\cal M} f$ .  The leading term in
Eq. (\ref{freqevol}) is the contribution of quadrupole radiation damping from
Eq. (\ref{Edotquad}).  The second term is the
contribution of dipole radiation damping 
[Eq. (\ref{Edotdipole})], where  $b = (5/48) {\cal
S}^2/\omega$.   Integrating Eq. (\ref{freqevol}) using only the leading term,
we
obtain approximate initial and final frequencies corresponding to an
observation period $T$ leading up to the innermost stable circular orbit
whose gravitational-wave
frequency is $f_{\rm ISCO} = ( 6^{3/2} \pi m_{BH})^{-1}$;  Eq.
(\ref{freqevol}) then yields the dipole contribution to the phase given by
$\Phi_D \approx 2.46 ({\cal S}^2/\omega\eta)[(1+16\eta
T/405m)^{7/8}-1]$.  For a $1.4 \,M_\odot$ neutron star and a
$10^4 \,M_\odot$ black hole, the frequency
at the ISCO is less than 0.5 Hz, while, with five million kilometer arm
lengths, the minimum time resolution for LISA is of order 30 seconds,
hence the minimum phase resolution for waves from such a black hole is
of order 100 radians.  Demanding that the accumulated phase offset due
to dipole radiation reaction be less than this resolution leads to the
crude bound on $\omega$ given by
\begin{equation}
\omega > 2 \times 10^4 \left ( {{\cal S} \over 0.3} \right )^2
              \left ( {100 \over \Delta\Phi_D } \right ) 
		\left ( {T \over {1 \,{\rm yr}}} \right )^{7/8}
		\left ( {{10^4 \, M_\odot} \over m_{BH} } \right )^{3/4}
\,.
\end{equation}


We confirm this crude estimate with detailed calculations using
matched filtering to estimate the parameters of such NS-MBH
inspiralling systems, and employing current proposals for the
instrumental noise curve for LISA together with an estimate for
confusion noise caused by a background of galactic white dwarf
binaries.  The results are displayed in Fig. 1.  Shown are the
bounds achievable for $1.4 M_\odot$
neutron stars inspiralling into black holes with
masses between 1000 and 200,000 $M_\odot$, and for integration times of
one year, 1/2 year and 1/4 year leading up to the innermost stable
orbit.  Notice that the bound decreases in almost inverse proportion
to  black hole
mass in agreement with our crude estimate; 
this is a result of the lower intrinsic frequencies of high mass
systems, and the consequent accumulation of fewer cycles in the
relevant integration time, leading to less ability to separate out the
dipole effects from quadrupole phasing.
Notice also that, for a given mass, the bound scales 
almost linearly with integration
time, again in agreement with our crude estimate.
Throughout, we
assume for concreteness a maximum (and thus) 
pessimistic value $s_{NS} = 0.2$ for 
the sensitivity of the neutron star, so that ${\cal S} = 0.3$.  If
$s_{NS}$ were as low as 0.1, the bounds would increase by $(0.4/0.3)^2
\approx 1.7$.  For $1.4 M_\odot$ white dwarf inspirals ($s_{WD} \ll s_{BH}$), 
the bounds would be larger
by $(0.5/0.3)^2 \approx 2.8$, provided that $M_{BH} > 2 \times 10^4
M_\odot$, so that the white dwarf reaches the ISCO without tidal disruption.
For $M_{BH} < 2 \times 10^4 M_\odot$, integration must be cut off as many as
a dozen days before the ISCO, with a consequent loss of sensitivity, so that
for a $1000 M_\odot$ black hole, white dwarf inspiral leads to almost the
same bound as does NS inspiral.

The remainder of the paper is devoted to details of the method and
calculations.  In Sec. \ref{sec:matched} we summarize the basic method
of parameter estimation using matched filtering; this method is based
on foundations laid for gravitational-wave detection by Finn and
Chernoff \cite{FinnChernoff} and Cutler and Flanagan \cite{CutlerFlanagan}, 
and applied to
specific parameter estimation problems for inspiralling systems by
Will and Poisson \cite{CW1,poissonwill,graviton}.  
In Sec. \ref{sec:bounds} we apply the
method specifically to observations by space interferometers.  First
we consider LISA.  Then we 
we apply the method to a hypothetical follow-on to the LISA mission
(which we call SuperLISA), whose sensitive band would lie between
that of LISA and the ground based interferometers.  The results are
discussed in Sec. \ref{sec:discussion}

\section{Estimating binary system parameters using matched filtering}
\label{sec:matched}

It is expected that laser interferometric observatories, whether ground-based or space-based, will 
endeavor to detect gravitational waves from binary inspirals using a
technique known as matched filtering \cite{3min}, wherein a
template consisting of a theoretical gravitational waveform is
compared to the detector output.  The template that is a match with an
actual signal present with the noise will show a strong correlation with
the detector output and thereby will ``filter'' a signal out of 
background noise.  The template is represented
by $h(t)$, related to the spatial components of the radiative metric
perturbation far from the source.  In the so-called ``restricted post-Newtonian
approximation'' for an inspiral orbit that is quasi-circular (that is, circular apart from the adiabatic 
decrease in separation), we approximate $h(t) \approx
\Re\{h_0(t)e^{i\Phi(t)}\}$, where $h_0(t)$ is a slowly varying wave amplitude;
it depends on the 
wave polarization, the
location of the source on the sky relative to the detector, 
the detector orientation,  and the distance
to the source; $\Phi(t)$ is the phasing of the wave, which is a
function of the evolving orbital frequency; and $\Re$ denotes the real
part.  
The wave amplitude $h_0(t)$ is considered
only to Newtonian order, {\it i.e.} given by the lowest-order,
quadrupole approximation,  because the process of matched-filtering is rather
insensitive to changes in the amplitude of the wave.  
Matched filtering is very sensitive to the phasing 
of the
wave, however, and thus $\Phi(t)$ is taken to the highest post-Newtonian order
possible.  The 
Fourier transform of $h(t)$ using the stationary phase
approximation is given by
\begin{equation}
\tilde{h}(f) = \left\{  \begin{array}{ll}
{\cal A} f^{-7/6} e^{i\Psi(f)} \,,  & 0<f<f_{\rm max} \\
0  \,, & f>f_{\rm max}
\end{array}  \right. 
\label{fourier}
\end{equation}
where $f_{\rm max}$ is the largest frequency for which the wave can be 
described by the restricted post-Newtonian approximation.  
For inspiral into black holes, this is often taken to correspond to waves 
emitted at the innermost stable orbit (ISCO) before the bodies 
plunge toward each other and merge.  In the
limit where one mass is much smaller than the other, this frequency is given by
\begin{equation}
f_{\rm max} = (6^{3/2} \pi m)^{-1}.
\label{eq.20}
\end{equation}
All the relevant factors related to the polarization of the wave, distance
to the source, and its position on the sky are
included in the amplitude ${\cal A}$.  
After averaging over all angles, we obtain
\begin{equation}
{\cal A} = \frac{1}{\sqrt{30} \pi^{2/3}} \frac {{\cal M}^{5/6}}{D_L} \,,
\label{amplitude}
\end{equation}
where  
$D_L$ is the ``luminosity'' 
distance of the source, 

Whereas the amplitude is calculated to Newtonian order, 
the phase  $\Psi(f)$ includes higher post-Newtonian 
corrections.  To Newtonian order in general 
relativity, $\Psi(f)$ is given by

\begin{equation}
\Psi(f) = 2 \pi f t_c - \phi_c - \pi/4 + \frac{3}{128} u^{-5/3} \,,
\label{psi}
\end{equation}
where $u=\pi {\cal M} f$ and $\phi_c$ is formally defined 
as the phase of the wave at the time of coalescence, $t_c$. 

In BD, for quasi-circular orbits, this formula is modified in two 
ways.  (1) To leading order, the 
modifications of quadrupole and monopole radiation can be subsumed in the simple replacement of 
the GR chirp mass $\cal M$ in Eqs. (\ref{amplitude}) and (\ref{psi})
by an effective chirp mass given by
\begin{equation}
{\cal M} \equiv \frac{\kappa^{3/5}}{G^{4/5}} \eta^{3/5} m \,,
\label{newchirpmass}
\end{equation}
where
\begin{eqnarray}
G &\equiv& 1 - \frac{s_1 + s_2 - 2 s_1 s_2}{2+\omega}, \nonumber \\
\kappa &\equiv& G^2 \left( 1 + \frac{1}{2(2+\omega)} +
\frac{[1-2(m_1s_2+m_2s_1)/m]^2}{12(2+\omega)} \right) \,,
\label{Gkappa}
\end{eqnarray}
and where  $s_1$ and $s_2$ are the sensitivities of the two  bodies.
(2)  Via the energy loss in Eq. (\ref{Edotdipole}), 
dipole radiation  introduces 
another
term in the phase $\Psi(f)$,
so that Eq. (\ref{psi}) now reads 
\begin{equation}
\Psi(f)_{BD} = 2 \pi f t_c - \phi_c - \pi/4 + \frac{3}{128} u^{-5/3} (
1-\frac{4}{7} b \eta^{2/5} u ^{-2/3} ) \,,
\label{psiBD}
\end{equation}
where the parameter $b$ is given by
\begin{equation}
b \equiv \frac{5}{48}\frac{\kappa^{-3/5} G^{4/5}}{2+\omega} {\cal S}^2 \,. 
\label{bparameter}
\end{equation}
As $\omega \to \infty$, $b \to 0$.   
For more details on the gravitational waveform $h(t)$ in BD, see
\cite{CW1}.  Because we assume {\it a priori} that $\omega > 3000$, we shall approximate 
$b \approx   ({5}/{48}) {\cal S}^2 /\omega $. 
We further add to the phasing formula the PN term and the 1.5 PN ``tail''
term of
general relativity; they serve the useful purpose in the matched
filter of breaking the degeneracies among the  various parameters
through their different frequency dependences.
While not strictly correct in BD, they are individually
valid up to corrections of order
$1/\omega$.  Thus we 
adopt the phasing \cite{CW1}
\begin{eqnarray}
\Psi(f) &=& 2 \pi f t_c - \phi_c - \frac{\pi}{4}  \nonumber \\
&& + \frac{3}{128}u^{-5/3}
\left[1-\frac{4}{7} b \eta^{2/5} u^{-2/3} 
+ \frac{20}{9}\left(
\frac{743}{336} + \frac{11}{4} \eta \right)\eta^{-2/5} u^{2/3} 
- 16 \pi \eta^{-3/5} u \right] \,,
\label{psifinal}
\end{eqnarray}
where the third and fourth terms inside the square brackets are the 
PN and 1.5PN terms, respectively.  Notice that, because $u \sim v^3$, the
dipole term is $O(1/v^2)$ compared to the quadrupole term, as expected.

By maximizing the correlation between a template waveform that depends on a set 
of parameters 
$\theta^a$  (for example, the chirp mass $\cal M$) and a measured signal, 
matched filtering provides 
a natural way to estimate the parameters of the signal and their errors 
(for discussion, see
\cite{FinnChernoff,CutlerFlanagan}).  With a given detector noise spectrum $S_n(f)$ one defines the inner product 
between two signals $h_1(t)$ and $h_2(t)$ by
\begin{equation}
(h_1|h_2) \equiv 2 \int_0^{\infty} \frac{ {\tilde{h}_1}^*\tilde{h}_2 +
{\tilde{h}_2}^*\tilde{h}_1 }{S_n(f)}df \,,
\label{innerproduct}
\end{equation}
where $\tilde{h}_1(f)$ and $\tilde{h}_2(f)$ are the Fourier transforms 
of the gravitational waveforms
$h(t)$.  The signal-to-noise ratio (SNR) for a 
given $h$ is given by
\begin{equation}
\rho[h] \equiv  (h|h)^{1/2} \,.
\label{rho}
\end{equation}
One then defines the ``Fisher information matrix'' $\Gamma_{ab}$ 
with components given by
\begin{equation}
\Gamma_{ab} \equiv \left( \frac{\partial h}{\partial\theta^a} \mid
\frac{\partial
h}{\partial\theta^b} \right) \,,
\label{fisher}
\end{equation}
An estimate of the rms error, $\Delta\theta^a$, in measuring  
the parameter $\theta^a$ can then be 
calculated, in the limit of large SNR, by
taking the square root of the diagonal elements of the inverse of the
Fisher matrix,
\begin{equation}
\Delta\theta^a = \sqrt{\Sigma^{aa}} \,, \qquad  \Sigma = \Gamma^{-1} \,.
\label{errors}
\end{equation}
The correlation coefficients between two parameters $\theta^a$ 
and  $\theta^b$  are given by
\begin{equation}
c^{ab} = \Sigma^{ab}/\sqrt{\Sigma^{aa}\Sigma^{bb}} \,.
\label{correlations}
\end{equation}

In our calculations, the parameters to be estimated will be 
$\phi_c$, $f_0 t_c$, $\ln {\cal M}$, $\ln \eta$, and $\tilde{b}$, 
where $\tilde{b}=b \eta^{2/5}$, 
and $f_0$ is a fiducial frequency characteristic 
of the detector noise spectrum.  We 
follow the method used in \cite{CW1}: combining Eqs. (\ref{fourier}) and
(\ref{psifinal}) and calculating the partial derivatives $\partial {\tilde h}
/\partial \theta^a$ for the five listed parameters, we construct the
Fisher information matrix using Eqs. (\ref{innerproduct}) and
(\ref{fisher}).  For simplicity, we set
$\tilde b$ equal to its nominal GR value of zero in all information
matrix expressions.  We then invert the information matrix and
evaluate the errors in the five parameters, along with the correlation
coefficients between $\cal M$, $\eta$ and $\tilde b$.  Since the
nominal value of $\tilde b$ is zero ($\omega = \infty$), the error on
it translates into a lower bound on $\omega$.  Results are calculated
for the canonical neutron star mass of $1.4 M_\odot$, and for 
various values of the black hole mass.

\section{Bounds on scalar-tensor gravity from space-based interferometers}
\label{sec:bounds}

\subsection{LISA-type interferometer}

We consider space-based interferometers of the proposed LISA type, with 
a sensitive bandwidth 
between $10^{-4}$ and $10^{-1}$ Hz, typical integration times up 
to one year, and an expected 
noise curve which can be expressed in terms of
an overall amplitude $S_0$, and a function of the ratio $x=f/f_0$:
\begin{equation}
S_n(f) = S_0 g(x) \,.
\end{equation}
Including the LISA instrumental noise curve and an estimate 
of ``confusion noise'' from a population 
of galactic white-dwarf binaries \cite{LISAphaseA,benderhils}, we adopt a noise curve given by
\begin{eqnarray}
S_0 &=& 4.2 \times 10^{-41} \, {\rm Hz}^{-1} \,, \nonumber \\
f_0 &=& 10^{-3} {\rm Hz} \,, \nonumber \\
g(x) &=& \sqrt{10} x^{-14/3} +1+x^2/1000 
 +313.5x^{-(6.398+3.518\log10 x)} \,.
\end{eqnarray}
The signal-to-noise ratio is given, from Eqs.  
(\ref{fourier}), (\ref{innerproduct}), and (\ref{rho}), by
\begin{equation}
\rho^2 = 4 |{\cal A}|^2 f_0^{-4/3} I(7) / S_0, 
\label{rhonew}
\end{equation}
where we define the integrals $I(q)$ by
\begin{equation}
I(q) = \int_{x_{\rm min}}^{x_{\rm max}} \frac{x^{-q/3}}{g(x)} dx \,,
\label{Iq}
\end{equation}
where  $ x_{\rm min} = f_{\rm min}/f_0$ and 
$ x_{\rm max} = f_{\rm max}/f_0$, corresponding 
to the minimum and maximum frequencies over which the detector will 
integrate .  In some calculations, the maximum 
value of $f_{\rm max}$  corresponds to 
radiation emitted at the innermost stable circular orbit of the 
black hole, with frequency $(6^{3/2} \pi m)^{-1}$, while in others, 
we consider the effect of terminating 
observations sooner than this final orbit.  
The frequency $ f_{\rm min} $ corresponds to the 
gravitational-wave frequency  observed a time $T$ earlier, 
where for LISA-type systems, 
$T \le $ one year.  Using the quadrupole approximation for radiation
damping [Eq. (\ref{freqevol}) with $b=0$], 
one can relate the frequencies of gravitational radiation at
the beginning and end of any time interval $T$ by the expression
\begin{equation}
u_i = u_f \left ( 1 + {256 \over 5} {T \over {\cal M}} u_f^{8/3} 
	\right )^{-3/8} \,.
\label{timeinterval}
\end{equation}

We will also wish to study the distance to which such sources can be
detected, in order to assess the likelihood of detection.
Combining Eqs. (\ref{amplitude}) and (\ref{rhonew}), we can relate the source 
luminosity distance $D_L$ to source 
masses and the SNR, 
\begin{eqnarray}
D_L &=& \sqrt{\frac{2}{15}} \frac{{\cal M}^{5/6}}{\rho} (\pi f_0)^{-2/3} \left( 
\frac{I(7)}{S_0} \right)^{1/2} \nonumber \\  
&=& 2.45 \, {\rm Gpc}  \left( \frac{m_{NS}}{1.4M_\odot} \right)^{1/2}
\left( \frac{m_{BH} }{10^4 M_\odot} \right)^{1/3}
\left( \frac{10}{\rho} \right)
\left( \frac{4.2 \times 10^{-41} }{S_0} \right)^{1/2}
\left( \frac{10^{-3} }{f_0} \right)^{2/3}
I(7)^{1/2} \,.
\label{lumdistance}
\end{eqnarray}
%


We first consider a $1.4 M_{\odot}$ 
NS in a quasi-circular inspiral into  a 
BH of mass ranging from $10^3 \, M_{\odot}$ to $2 \times 10^5 \,
M_{\odot}$ and an integration
time on LISA of one year leading up to the innermost stable orbit.  We
assume a SNR of 10.
The parameter errors found are shown in Table~\ref{tab:table1}. 
For MBH masses less than about 70,000 $M_\odot$, the bound on $\omega$
exceeds the current solar-system value of 3000, and for low mass
black holes ($10^3 M_\odot$), the bound could be 80 times larger.  We
then determine the bounds on $\omega$ for the same range of masses,
and for orbits terminating at the ISCO,
but for shorter integration times, such as 1/2 year or 1/4 year.  The
results are plotted in Fig. 1.  Also shown in Fig. 1 
are the approximate distances to
which such systems can be observed.  As
expected, the bounds weaken almost in direct proportion to integration
time.   Figure 2 illustrates the effect of failing to observe the
final orbits leading to the ISCO: shown for various BH masses are the
bounds for one year of integration terminating a given number of days
before the ISCO.  The rapid fall-off of the bound with time before the ISCO
indicates the importance of seeing the highest frequency, most relativistic
orbits.  
For sources at a given redshift (corresponding to a
given $D_L$), the SNR can be calculated from 
Eq. (\ref{lumdistance}), and this
used to determine the relevant bound on $\omega$.  The results for $Z=0.01,
\, 0.03 $ and $0.05$ are
plotted in Fig. 3.  To convert from $D_L$ to $Z$, we use the formula
for a matter dominated, spatially flat cosmology
$D_L=(2/H_0)(1+Z)[1-(1+Z)^{-1/2}]$ with a value of the Hubble
parameter $H_0 = 50 \,{\rm km} \, {\rm s}^{-1} \, {\rm Mpc}^{-1}$.  
The signal-to-noise ratios for several illustrative cases are shown in the
figure.


In principle, white-dwarf inspirals should give stronger bounds on $\omega$
because their sensitivities are much smaller than that of black holes.
However, for small enough black hole mass, tidal disruption will require 
terminating the matched filter before the ISCO, with a consequent loss of
accuracy in bounding $\omega$, as indicated in Fig. 2.
We take, as a crude measure of the termination of integration for a white
dwarf inspiral, the Roche radius $R \approx d (m_{BH}/m_{WD})^{1/3}$, where
$d$ is the radius of the white dwarf, or equivalently, the place where the
gravitational-wave frequency is given by $\pi f = m_{BH}/r^3 = m_{WD}/d^3$.
Using the approximation, for a non-relativistic WD equation of state, $d
\approx 6 \times 10^3 M_\odot (m_{WD}/M_\odot)^{-1/3}$, and substituting
into Eq. (\ref{timeinterval}), we find an approximate cut-off time
before the ISCO given by
\begin{equation}
T_{\rm cut-off} \approx 4 \, {\rm days} \, 
	\left ( {{m_{BH}} \over {10^3 M_\odot}} \right )^{-2/3}
	\left ( {{m_{WD}} \over {1.4 M_\odot}} \right )^{-11/3}
        \left \{ 1 - {1 \over 4000} 
	\left ( {{m_{BH}} \over {10^3 M_\odot}} \right )^{8/3}
	\left ( {{m_{WD}} \over {1.4 M_\odot}} \right )^{8/3}
	\right \} \,.
\end{equation}
When $T$ becomes negative (for $m_{BH} > 2.2 \times 10^4 M_\odot$, for a
$1.4 M_\odot$ WD), the cut-off
is the ISCO.  Using the Fisher information matrix, we calculate the bounds
on $\omega$ for one year of integration of white dwarf inspirals down to the
appropriate cut-off time, with a SNR of 10.  
For $m_{BH} > 2.2 \times 10^4 M_\odot$, the
bounds are 2.8 times better than for neutron stars of the same mass, as
expected from the difference in sensitivities, but for $m_{BH} < 2.2 \times
10^4 M_\odot$, the improvement gradually decreases because of the loss of
orbits near the ISCO.  For a $10^3 M_\odot$ BH, the bound is only 14 percent
better than the corresponding neutron-star bound.

\subsection{Follow-up Missions to LISA}

Concepts for a future 
advanced space-based interferometer have been developed, based on the
LISA project, which, for want of an official name, we will dub
``SuperLISA''.  The motivation for this mission is to detect gravitational
waves with a peak sensitivity between the ranges of LISA and the
ground-based detectors, say around $10^{-1}$ Hz.  Here, it is hoped, there
will exist a window free of astrophysically generated sources of background
gravitational waves, such as the white dwarf binaries that contribute to
noise in LISA below $10^{-2}$ Hz.  With such a window, it is hoped that
cosmological backgrounds of gravitational waves may be 
detectable.   

A
hypothetical noise curve for this advanced detector is given by $S_n(f) =
S_0 g(x)$ with $x=f/f_0$, where  \cite{SuperLISA}

\begin{eqnarray}
S_0 &=& 2.1 \times 10^{-47} \, {\rm Hz}^{-1} \,, \nonumber \\
f_0 &=& 10^{-1} {\rm Hz} \,, \nonumber \\
g(x) &=&  0.579 x^{-4} +1+ 0.064 x^2 \,.
\label{superlisanoise}
\end{eqnarray}

With this sensitivity, SuperLISA should be able to detect inspirals of
intermediate mass systems ($100 - 10^5 M_\odot$) to cosmological distances
with large SNR.  
Applying the method of the previous subsection to the noise curve of Eq.
(\ref{superlisanoise}), we find that 
detection of these signals from NS-MBH inspirals may allow SuperLISA to
place even more dramatic bounds on scalar-tensor theory than LISA. 

We assume integration times of one year terminating at the ISCO 
and a neutron star sensitivity of
$s_{NS} = 0.2$, and  consider sources at fixed redshifts of $Z = 1 ,\, 5 ,\,
{\rm and} \, 10$.
The bounds obtainable
from a $1.4 M_{\odot}$ NS inspiralling to black holes of a few hundreds of
solar masses exceed tens of millions,  
and are dramatically better than the solar-system value
for essentially all BH masses between $100 \,M_\odot$ and $10^5
M_\odot$ (Fig. 4).
For sources at redshift $Z = 1$ and $Z= 5$, the SNR is
well above 10.  
The drop in the bound on $\omega$
for $M_{BH} < 150 M_\odot$ is caused by the fact that the initial
frequency (one year before the ISCO) for such low-mass systems is 
already around $10^{-1}$ Hz, where the instrument has its highest
sensitivity; most of the data in these cases is being taken against a
rapidly increasing background of instrumental noise (mostly caused by lack
of time resolution related to the instrument's arm lengths), hence the
bound weakens.


\section{Conclusions}
\label{sec:discussion}

We have found that 
future observations of inspirals of neutron stars into massive black holes
by space-based laser interferometric detectors such as LISA may 
place significant bounds on the scalar-tensor coupling
parameter $\omega$.  For inspirals into black holes as low as $10^3
M_{\odot}$, the bound could be as large as 240,000 (for a SNR of 10), 
80 times larger than
current solar-system bounds.  The bound achievable decreases with increasing
black hole mass, for a given SNR.  A follow-up space
interferometer, with a sensitive band at intermediate frequencies between
those of LISA and the ground based interferometers could produce bounds in
the tens of millions.

The bound on $\omega$ is a strong function of the integration time.  
For integration times shorter than the canonical one year,
$\omega$ decreases almost in direct proportion, and if the
incoming wave is not integrated up to the last stable orbit, then the
possible bound on $\omega$ drops off sharply. 
For neutron-star sensitivities smaller than 0.2, all quoted bounds increase
by the factor $[(0.5 - s_{NS})/0.3]^2$; 

Two important issues have not been addressed.

The first is our restriction to quasi-circular orbits.   This
approximation is reasonable for stellar mass inspirals of compact objects in
galaxies, which are expected to be detectable by ground-based
interferometers.  These are systems 
where gravitational radiation damping has had sufficient time to
circularize any pre-existing, two-body eccentric orbit.  However, in dense
galactic nuclei or in dense globular clusters containing massive black
holes, compact objects are expected to be injected into highly eccentric
orbits via interactions with a cloud of objects surrounding the hole, and to
suffer frequent perturbations by these bodies during apholion passage.

Eccentricity in and of itself should not have a strong effect on the bounds
we have inferred.
A preliminary estimate of the effect of small
eccentricity on wave detection shows a small drop in
SNR for an inspiral with a small eccentricity.  For
eccentricities of $e = 0.1$ or less, the drop in SNR be four
percent or less
if a quasi-circular inspiral template is used to match against an
eccentric waveform.  For eccentricities of 0.2 to 0.3, the drop in
SNR will increase to at least 10\% and may be as large as 35\% for some
systems.  This drop in SNR will be accompanied by a similar drop in
detection rates and in the bounds on $\omega$.  This drop is not too
large, however, and can be made even less by using an eccentric waveform
template (with appropriate BD terms) 
to match the detected gravitational wave.  On the other hand, frequent,
essentially random perturbations of the orbit of the compact object by
surrounding stars will make parameter estimation very difficult, even in
general relativity.  How important this effect will be, and whether there
exist good data analysis techniques to handle it, is a subject of active
investigation.

The second issue is event rate for inspirals into the intermediate mass
black holes ($100 - 10^5 M_\odot$) that give the best bounds on
$\omega$.
While enough is known and speculated about the
rate of occurrence of $10^6 M_{\odot}$ or $10^7 M_{\odot}$ binary BHs in the
centers of galaxies to make such systems promising sources for LISA, 
little to
nothing is known about black holes of mass $10^4 M_{\odot}$ or
less.  Until some estimate can be made of how $10^2 M_{\odot} - 10^4
M_{\odot}$ BH are distributed in the Universe, it will also be uncertain
whether or not neutron stars can be expected to exist around these black
holes, and how these neutron stars are expected to behave as they
inspiral.  Nevertheless, there has been recent speculation, based on
numerical simulations, modeling and some observation, that runaway growth of
intermediate mass black holes may occur in globular clusters or in
young compact star clusters near AGNs
\cite{miller,missinglink}.   Another mechanism that has been discussed is
the gravitational collapse of the first stars, assumed to be 
very massive objects $10^2 - 10^5 M_\odot$
\cite{schneider}.  

It is worth pointing out that testing fundamental theory does not require
the same event rate that a viable gravitational-wave astronomy does.
A rare occurrence of a source with just the right characteristics can
provide a strong test of scalar-tensor gravity.  By comparison, while 35
years of pulsar astronomy boasts an average pulsar discovery rate of about
3 per month (for a total of over 
1000 pulsars), it took only one, PSR 1913+16, to
test the general relativistic quadrupole for gravitational radiation
damping.  Still, an estimate of the rate of compact body inspiral into
intermediate mass black holes would be desirable, in order to evaluate
how truly feasible such tests of scalar-tensor gravity using space
interferometers might be.

\section*{Acknowledgments}

We are grateful to Achamveedu Gopakumar, Matt Visser, and Sterl Phinney
for useful discussions.  This work is supported in part by the National
Science Foundation Grant No. PHY 96-00049 and PHY 00-96522, and the National
Aeronautics and Space Administration Grant No. NAG 5-10186.

\begin{figure}[t]
\begin{center}
\leavevmode
\psfig{figure=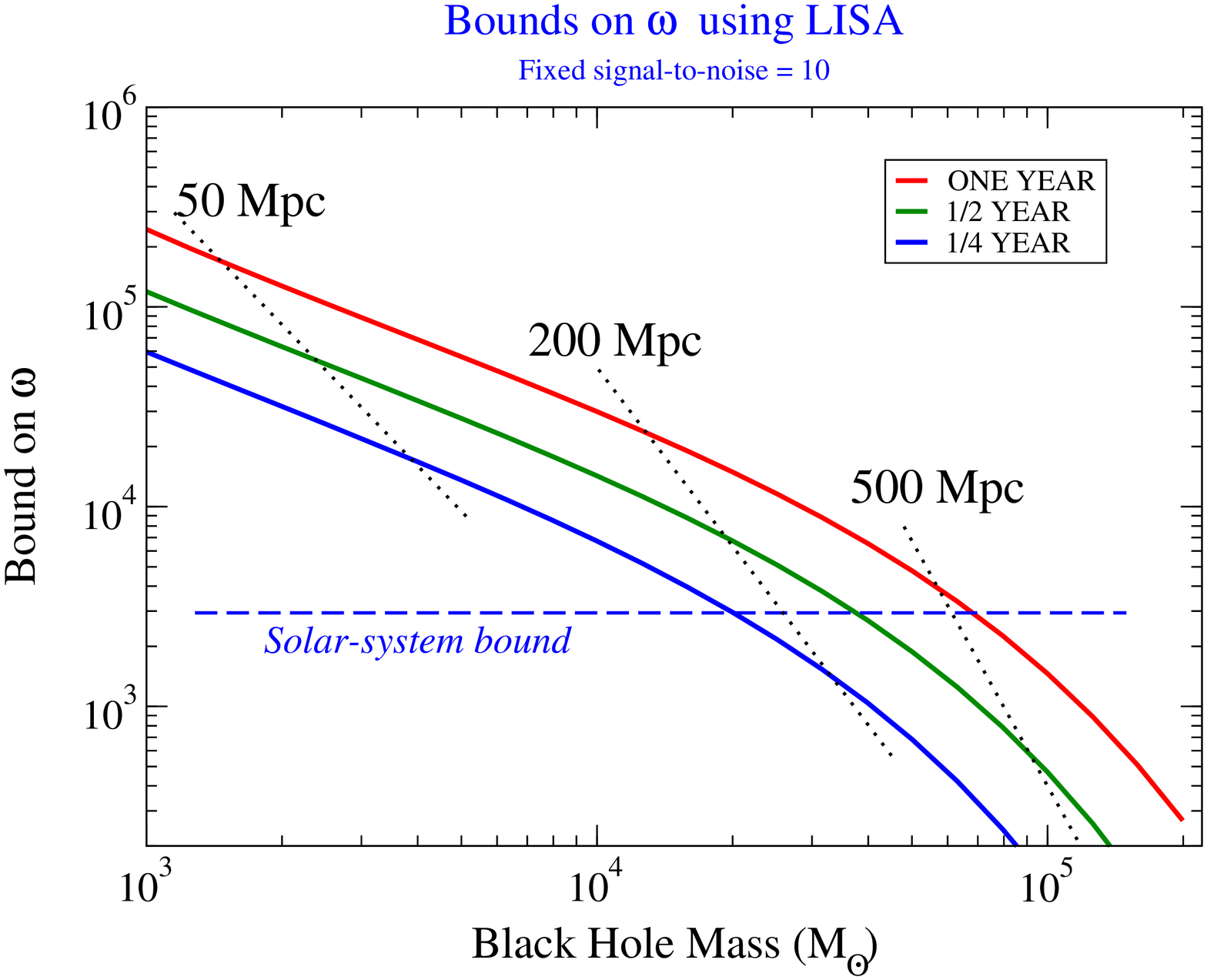,angle=270,height=5.0in}
\end{center}
\caption{Possible bounds on $\omega$ vs. BH mass:  Three different
integration times for LISA are shown, one, 1/2, and 1/4 year.
A $1.4 M_\odot$ neutron star with $s_{NS}=0.2$ and a SNR of 10 are assumed.   
Also shown are approximate
distances to the sources, along with the current solar-system bound 
on $\omega$ of 3000.}
\label{fig:figure1}
\end{figure}

\begin{figure}[t]
\begin{center}
\leavevmode
\psfig{figure=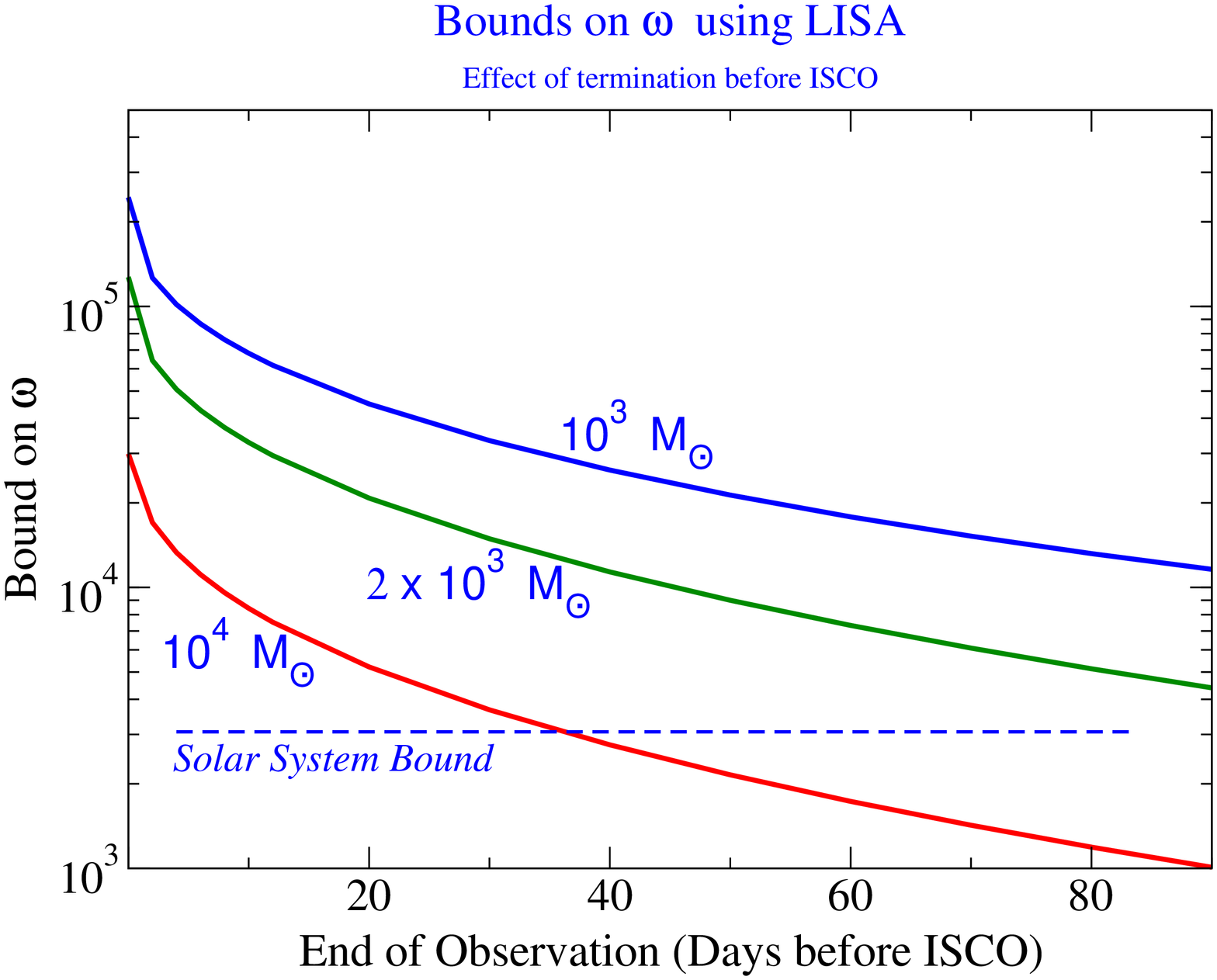,angle=270,height=5.0in}
\end{center}
\label{fig:figure2}
\caption{Effect of endpoint of integration, on bound on
$\omega$, in days before the ISCO.   
Bounds achievable on $\omega$ drop off rapidly if signal is not
integrated up to last stable orbit.  }
\end{figure}

\begin{figure}[t]
\begin{center}
\leavevmode
\psfig{figure=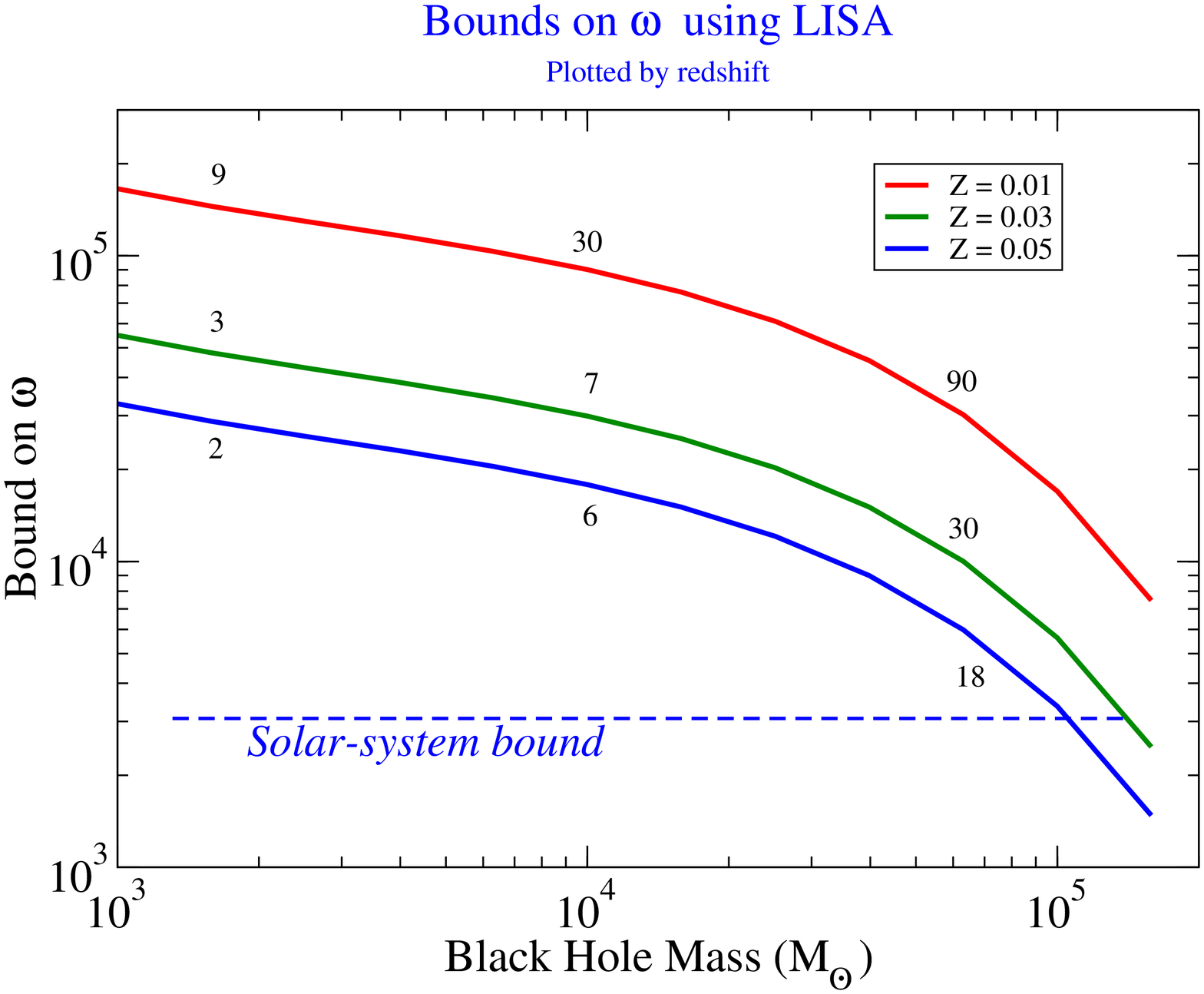,angle=270,height=5.0in}
\end{center}
\label{fig:figure3}
\caption{Possible bounds on $\omega$ vs. BH mass as a function of source
redshift.  Numbers above curves indicate approximate SNR for
detection.  }
\end{figure}

\begin{figure}[t]
\begin{center}
\leavevmode
\psfig{figure=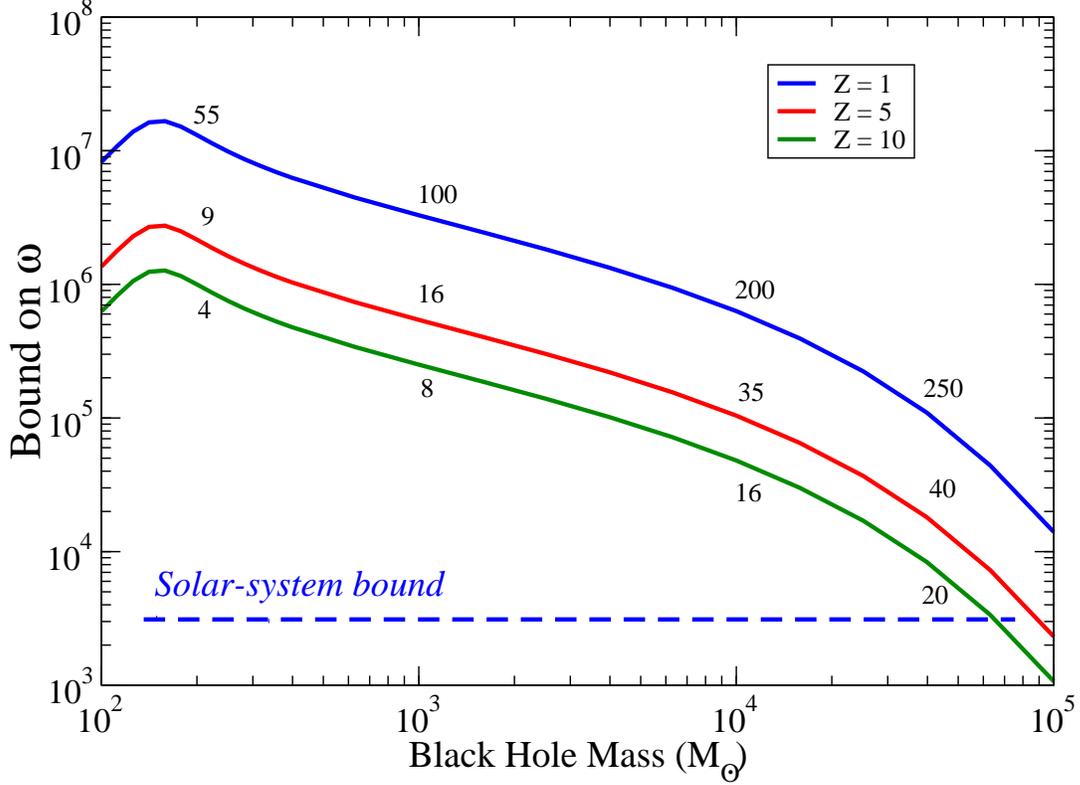,angle=270,height=5.0in}
\end{center}
\bigskip
\label{fig:figure4}
\caption{Bounds on $\omega$ vs. BH mass for SuperLISA and sources 
at redshifts of 
$Z=$ 1, 5 and 10.  
A $1.4 M_\odot$ neutron star with $s_{NS}=0.2$ is assumed.
Numbers above curves indicate approximate SNR for
detection.  Drop in sensitivity for $M_{BH} < 150 M_\odot$ is because
GW frequencies are above point of SuperLISA's maximum sensitivity.}
\end{figure}

\begin{table}[t]
\leavevmode
\begin{tabular}{rrrrrrrrr}
$m_{BH} $&$\Delta t_c$&&$\Delta {\cal M}/{\cal M}$&$\Delta\eta / \eta$&Bound
& \\ 
$(M_\odot)$&(s)&$\Delta \Phi_c$&\hfil (\%)\hfil &\hfil (\%)\hfil &\hfil
on $\omega$\hfil  
& $c^{\eta {\cal M}}$ & $c^{{\cal M} \tilde{b}}$ & $c^{\eta\tilde{b}}$ \\ \hline

1000&	2.39&	12.3&	.000192&.0727&	244549&	.885&	-.996&	-.919
\\

5000&	3.80&	11.7&	.000544&.0252&	56326&	.970&	-.998& -.953
\\

10000&	5.32&	12.9&	.000768&.0180&	29906&	.978&	-.997&	-.961
\\

50000&	26.54&	32.0&	.00243&	.0145&	4780&	.989&	-.998&	-.978
\\

100000&	103.07&	84.8&	.00610&	.0219&	1458&	.993&	-.999&	-.986

\end{tabular}

\caption{Estimated parameter errors for $1.4 \, M_\odot$ NS-MBH 
systems:  SNR = 10, integration 
time
is one year prior to the ISCO, neutron star sensitivity $s_{NS} = 0.2$. 
Error in coalescence time $t_c$ is in seconds; error in coalescence
phase is in radians.
Errors in chirp mass $\cal M$ and reduced mass parameter $\eta$ are
in percent.
Dimensionless correlation
coefficients $c^{{\cal M} \eta}, c^{{\cal M} \tilde{b}}$, 
and $c^{\eta \tilde{b}}$ are 
also given.} 
\label{tab:table1} 
\end{table}


\begin{references}
\bibitem{caveat}
The exception to this statement is a class of generalized scalar-tensor
theories in which the scalar field exhibits an instability to ``spontaneous
scalarization'' in the strong-field interiors of neutron stars, thereby
enhancing the difference in sensitivities between even quite similar neutron
stars, such as in the binary pulsar.  For details, see
\cite{damouresposito}.
\end{references}
\end{document}